\documentclass[journal]{IEEEtran}

\usepackage[dvips]{color}
\usepackage{graphicx}
\usepackage{amsmath}
\begin{document}
%
\title{Physical-Layer Security with Multiuser Scheduling in Cognitive Radio Networks}
%
%
%
\markboth{IEEE Transactions on Communications (accepted to appear)}%
{Yulong Zou \MakeLowercase{\textit{et al.}}: Physical-Layer Security with Multiuser Scheduling in Cognitive Radio Networks}

\author{Yulong~Zou,~\IEEEmembership{Senior Member,~IEEE,}
        Xianbin~Wang,~\IEEEmembership{Senior Member,~IEEE,}
        and Weiming~Shen,~\IEEEmembership{Fellow,~IEEE}

\thanks{Manuscript received April 2, 2013; revised July 25,
2013 and September 19, 2013; accepted November 1, 2013. The editor coordinating the review of this paper and approving it for publication was Prof. Husheng Li.}
\thanks{This work was partially supported by the National Natural Science Foundation of China (Grant Nos. 61302104, 61271240), the Scientific Research Foundation of Nanjing University of Posts and Telecommunications (Grant No. NY213014), and the Auto21 Network of Centre of Excellence, Canada.}
\thanks{Y. Zou is with the School of Telecommunications and Information Engineering, Nanjing University
of Posts and Telecommunications, Nanjing 210003, China. He was with the Electrical and Computer Engineering Department, University of
Western Ontario, London, Ontario, Canada, N6A 5B9. (E-mail:
yulong.zou@gmail.com).}
\thanks{X. Wang is with the Electrical and Computer Engineering Department, University of Western
Ontario, London, Ontario, Canada, N6A 5B9.}
\thanks{W. Shen is with the College of Electronics and Information Engineering, Tongji University, Shanghai, China, and with National Research Council, Ottawa, Canada.}

}

\maketitle

\begin{abstract}
In this paper, we consider a cognitive radio network that consists of one cognitive base station (CBS) and multiple cognitive users (CUs)
in the presence of multiple eavesdroppers, where CUs transmit their
data packets to CBS under a primary user's quality of service (QoS)
constraint while the eavesdroppers attempt to intercept the
cognitive transmissions from CUs to CBS. We investigate the
physical-layer security against eavesdropping attacks in the cognitive radio network and propose the user scheduling scheme to achieve multiuser diversity for improving the security level of cognitive transmissions with a primary QoS constraint. Specifically, a cognitive user (CU) that satisfies the primary QoS requirement and maximizes the achievable secrecy rate of cognitive transmissions is scheduled to transmit its data packet. For the comparison purpose, we also examine the traditional multiuser scheduling and the artificial noise schemes. We analyze the achievable secrecy rate and intercept
probability of the traditional and proposed multiuser scheduling
schemes as well as the artificial noise scheme in Rayleigh fading
environments. Numerical results show that given a primary QoS constraint, the proposed multiuser scheduling scheme generally outperforms the traditional multiuser scheduling and the artificial noise schemes in terms of the achievable secrecy rate and intercept probability. In addition, we derive the diversity order of the proposed multiuser scheduling scheme through an asymptotic intercept probability analysis and prove that the full diversity is obtained by using the proposed multiuser scheduling.

\end{abstract}


\IEEEpeerreviewmaketitle

\section{Introduction}
%
%
%
%
\IEEEPARstart {C}{ognitive} radio has been recognized as a promising
technology that enables unlicensed secondary users (also known as cognitive users) to dynamically access the licensed spectrum that is
assigned to but not being used by primary users (PUs) [1]-[3]. In
cognitive radio, PUs have a higher priority in accessing the
licensed spectrum than cognitive users (CUs). At present, there are
two main cognitive radio paradigms: 1) overlay cognitive radio [4],
[5], in which CUs first identify the white space of the licensed
spectrum (called spectrum hole) through spectrum sensing and then
utilize the detected spectrum hole for data transmission; and 2)
underlay cognitive radio [6], [7], in which CUs and PUs are allowed
to simultaneously access the licensed spectrum as long as the
quality of service (QoS) of PUs is not affected. Recent years have
witnessed an increasing interest on the cognitive radio topic which
has been studied extensively from different perspectives (see [8]
and references therein).

Cognitive radio security has been attracting continuously growing
attention, due to the open and dynamic nature of cognitive radio
architecture where various unknown wireless devices are allowed to opportunistically access the licensed spectrum, which makes cognitive radio systems vulnerable to malicious attacks. Recently, considerable research efforts have been devoted to the cognitive radio security
against the primary user emulation (PUE) attack [9] and
denial-of-service (DoS) attack [10]. More specifically, a PUE
attacker attempts to emulate a primary user (PU) and transmits
signals with the same characteristics as the PU, which constantly
prevents CUs from accessing the spectrum. In contrast, a DoS
attacker intentionally transmits any signals (not necessarily
emulating the PU's signal characteristics) to generate interference
and disrupt the legitimate user communications. It needs to be
pointed out that both the PUE and DoS attackers transmit harmful
active signals which can be detected by authorized users and then
prevented with certain strategies. In addition to the active PUE and
DoS attackers, an eavesdropper is a passive attacker that attempts
to intercept the legitimate transmissions, which is typically
undetectable since the eavesdropper keeps silent without
transmitting any active signals. Traditionally, the cryptographic
techniques relying on secret keys have been employed to protect the
communication confidentiality against eavesdropping attacks, which,
however, increases the computational and communication overheads and
introduces additional system complexity for the secret key
distribution and management.

As an alternative, physical-layer security is now emerging as a new
secure communication means to defend against eavesdroppers by
exploiting the physical characteristics of wireless channels. This
work was pioneered by Shannon in [11] and further extended by Wyner
in [12], where a so-called \emph{secrecy capacity} is developed from
an information-theoretical prospective and shown as the difference
between the capacity of the channel from source to destination
(referred to as main link) and that of the channel from source to
eavesdropper (called wiretap link). It was proved in [13] that if
the capacity of the main link is less than that of the wiretap link,
the eavesdropper will succeed in decoding the source signal and an
intercept event occurs in this case. To improve the physical-layer
security of wireless transmissions, some recent work was proposed by
exploiting the multiple-input multiple-output (MIMO) [14]-[16] and
cooperative relays [17], [18]. It was shown that the secrecy
capacity significantly increases through the use of MIMO and user
cooperation techniques. Notice that the aforementioned work [14]-[18] addresses the traditional non-cognitive radio networks and the
physical-layer security against eavesdropping attacks is rarely
investigated for cognitive radio networks. Compared to traditional
wireless networks, there are some unique challenges to be addressed
for the physical-layer security in cognitive radio networks,
e.g., the PU's QoS protection issue and the mitigation of mutual interference between PUs and CUs. More specifically, when applying the MIMO and user cooperation techniques for physical-layer security in cognitive radio networks, we need to consider how to avoid causing additional harmful interference to PUs. This requires certain modification of the conventional MIMO and user cooperation mechanisms for protecting the PUs' QoS while maximizing the cognitive radio security.

In this paper, we explore the physical-layer security in a cognitive
radio network consisting of one cognitive base station (CBS) and multiple cognitive users (CUs), where multiple eavesdroppers are assumed to intercept the cognitive transmissions from CUs to CBS. The main contributions of this paper are summarized as follows. Firstly, we propose the multiuser scheduling scheme to achieve multiuser
diversity for improving the cognitive transmission security against
eavesdropping attacks, in which a cognitive user (CU) that maximizes
the achievable secrecy rate of cognitive transmissions is scheduled
for data transmission under the PU's QoS constraint. Secondly, we examine the traditional multiuser scheduling [19], [20] and the artificial noise scheme [21], [22] for the purpose of comparison with the proposed multiuser scheduling. Also, we evaluate the security performance of the traditional and proposed multiuser scheduling schemes as well as the artificial noise scheme in terms of the achievable secrecy rate and intercept probability. Finally, we conduct the diversity order analysis and show that the proposed multiuser scheduling scheme achieves the diversity order of $M$, where $M$ represents the number of CUs.

The remainder of this paper is organized as follows. Section II
presents the system model of cognitive transmissions in the presence
of multiple eavesdroppers. In Section III, we propose the multiuser
scheduling scheme for cognitive transmissions to defend against
eavesdropping attacks and analyze the achievable secrecy rate in
Rayleigh fading channels. For the comparison purpose, we also present the traditional multiuser scheduling and the artificial noise schemes and analyze their achievable secrecy rates. Section IV conducts the intercept probability analysis of the multiuser scheduling and the artificial noise schemes and shows the security advantage of the proposed multiuser scheduling approach over the artificial noise scheme. Next, Section V presents the diversity order analysis and proves the full diversity achieved by the proposed multiuser scheduling scheme. Finally, we provide some concluding remarks in Section VI.

\section{System Model}
\begin{figure}
  \centering
  {\includegraphics[scale=0.55]{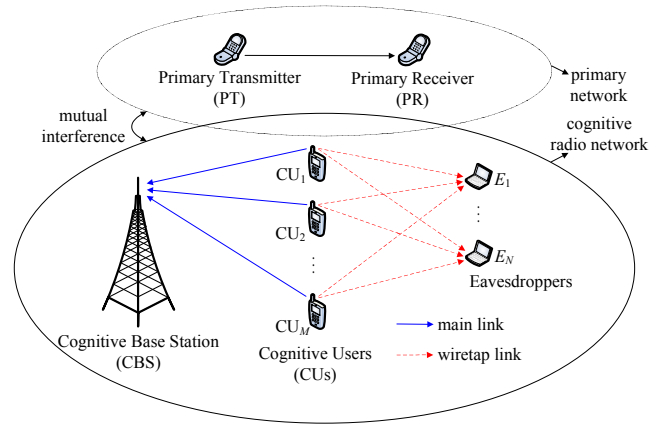}\\
  \caption{{Coexistence of a
  primary network consisting of one primary transmitter
  (PT) and one primary receiver (PR) with a cognitive radio
   network consisting of one cognitive base station (CBS) and $M$ cognitive
   users (CUs) in the presence of $N$ eavesdroppers.}}\label{Fig1}}
\end{figure}

In cognitive radio networks, CUs are allowed to access the licensed
spectrum that is assigned to PUs and mutual interference typically
exists between CUs and PUs. As shown in Fig. 1, we consider a
cognitive radio network that consists of one CBS and $M$ CUs, e.g.,
an IEEE 802.22 wireless regional area network (WRAN) [2] where a CBS
serves multiple customer premise equipments (CPEs) which are also
known as CUs. In Fig. 1, $N$ eavesdroppers are assumed to intercept
the cognitive transmissions from CUs to CBS, where all nodes are
equipped with single antenna and the solid and dash lines
represent the main and wiretap links, respectively. For notational
convenience, $M$ CUs and $N$ eavesdroppers are denoted by
${\cal{U}}=\{U_i|i=1,2,\cdots,M\}$ and
${\cal{E}}=\{{{E}}_j|j=1,2,\cdots,N\}$, respectively.
{Throughout this paper, we assume the
underlay cognitive radio, i.e., when PT is transmitting to PR over a
spectrum band, a CU is also allowed to transmit its data to CBS over
the same band at the same time as long as the PU's QoS is not
affected. In order to protect the PU's QoS, the interference
received at PR shall be guaranteed not to exceed the maximum
tolerable level denoted by $I$ through limiting the transmit power
of $\textrm{CU}_i$ ($P_i$), yielding}
\begin{equation}\label{equa1}
{P_i} = \frac{I}{{|{h_{ip}}{|^2}}},
\end{equation}
where ${h_{ip}}$ represents the fading coefficient of the channel
from $\textrm{CU}_i$ to PR. As shown in Eq. (1), for the PU's QoS protection, the transmit power of a cognitive user becomes a random variable, which leads to some new challenges in the resultant secrecy rate and intercept probability analysis. Due to the mutual interference between primary and cognitive transmissions [3], the interference will also be received at CBS from PT. As discussed in [23] and [24], the interference caused by PT to CBS can be modeled as a Gaussian random process, since the primary signal is typically processed at CBS with non-coherent detection. Moreover, the thermal noise follows a complex Gaussian distribution, which makes the interference plus noise at CBS become a complex normal random variable with zero mean and variance $N_b$ as denoted by ${\mathcal{CN}}(0,N_b)$, where subscript $b$ denotes CBS. Similarly, the interference and noise received at an eavesdropper $E_j$ can also be modeled as a complex normal random variable with zero mean and variance $N_{e_j}$ where $E_j \in {\cal{E}}$. In the cognitive radio network, CBS is regarded as a centralized controller and its associated CUs access the licensed spectrum using an orthogonal multiple access method such as the orthogonal frequency division multiple access (OFDMA) as specified in IEEE 802.22 network [2]. Considering that $\textrm{CU}_i$ transmits its signal $x_i$ with power $\frac{I}{{|{h_{ip}}{|^2}}}$, the received signal at CBS is given by
\begin{equation}\label{equa2}
{y_b} = \sqrt {\frac{I}{{|{h_{ip}}{|^2}}}} {h_{ib}}{x_i} + {n_b},
\end{equation}
where ${h_{ib}}$ represents the fading coefficient of channel from
$\textrm{CU}_i$ to CBS and $n_b \sim {\mathcal{CN}}(0,N_b)$ denotes
the interference and thermal noise received at CBS. Meanwhile, due
to the broadcast nature of wireless medium, the eavesdroppers can
overhear the transmission from $\textrm{CU}_i$ to CBS. Following the
physical-layer security literature (see [15]-[18] and references
therein), we assume that the eavesdroppers are aware of all the
parameters of the cognitive transmission from $\textrm{CU}_i$ to
CBS including the carrier frequency, spectrum bandwidth, coding and
modulation scheme, encryption algorithm, and so on, except that the
source signal $x_i$ is confidential. Hence, the received signal at
eavesdropper $E_j$ is given by
\begin{equation}\label{equa3} {y_{e_j}} = \sqrt
{\frac{I}{{|{h_{ip}}{|^2}}}} {h_{ie_j}}{x_i} + {n_{e_j}},
\end{equation}
where ${h_{ie_j}}$ represents the fading coefficient of the channel
from $\textrm{CU}_i$ to $E_j$ and $n_{e_j} \sim
{\mathcal{CN}}(0,N_{e_j})$ denotes the interference and thermal
noise received at eavesdropper $E_j$. In this paper, we assume that
$N$ eavesdroppers independently perform their tasks to intercept the
cognitive transmission. If none of $N$ eavesdroppers succeeds in
decoding the source signal, the cognitive transmission from
$\textrm{CU}_i$ to CBS is secure; otherwise, the cognitive
transmission is not secure and an intercept event is considered to
occur in this case. Both the main and wiretap links as shown in Fig.
1 are modeled as Rayleigh fading channels, i.e., $|h_{ip}|^2$,
$|h_{ib}|^2$ and $|h_{ie_j}|^2$ follow exponential distributions
with means $\sigma_{ip}^2$, $\sigma_{ib}^2$ and $\sigma_{ie_j}^2$,
respectively. Although only Rayleigh fading channels are considered
throughout this paper, similar performance analysis and results can
be obtained for other wireless fading models (e.g., Nakagami fading)
with path loss and shadowing. In addition, we assume that the
global channel state information (CSI) of both the main links and
wiretap links is available in performing the multiuser scheduling,
which is a common assumption in the physical-layer security
literature (e.g., [25] and [26]). Notice that studying the case of
the global CSI available can provide a theoretical performance upper
bound on the achievable secrecy rate of cognitive transmissions. We
will also examine the traditional multiuser scheduling scheme that
does not need the eavesdroppers' CSI.

\section{Proposed Multiuser Scheduling and Achievable Secrecy Rate Analysis}
In this section, we propose the multiuser scheduling scheme to
improve the physical-layer security of cognitive transmission and
analyze its achievable secrecy rate.
{For the comparison purpose, the
traditional multiuser scheduling and the artificial noise approaches
are also presented as benchmark schemes.} Numerical secrecy rate
results are provided to show the physical-layer security improvement
by exploiting the multiuser scheduling.

\subsection{Proposed Multiuser Scheduling Scheme}
This subsection presents the multiuser scheduling scheme to defend
against eavesdropping attacks. Given a spectrum band available for
the cognitive transmission, a CU among $M$ CUs should be selected
and scheduled for its data transmission to CBS. Without loss of
generality, consider that $\textrm{CU}_i$ is scheduled to transmit
its signal $x_i$ to CBS. Assuming the optimal Gaussian codebook used
at $\textrm{CU}_i$ and using Eq. (2), we can obtain the achievable
rate at CBS as
\begin{equation}\label{equa4}
{R_b}(i) = {\log _2}(1 +
\frac{{I|{h_{ib}}{|^2}}}{{|{h_{ip}}{|^2}{N_b}}}),
\end{equation}
where $U_i \in {\cal{U}}$. Meanwhile, from Eq. (3), the achievable
rate at eavesdropper $E_j$ from $\textrm{CU}_i$ is given by
\begin{equation}\label{equa5} {R_{e_j}}(i) = {\log _2}(1 +
\frac{{I|{h_{ie_j}}{|^2}}}{{|{h_{ip}}{|^2}{N_{e_j}}}}),
\end{equation}
where $E_j \in {\cal{E}}$. Considering that $N$ eavesdroppers
independently perform their interception tasks, the overall rate of
the wiretap links is the maximum of individual rates achieved at $N$
eavesdroppers. Thus, the overall rate ${R_{e}}(i)$ is the highest
one among ${R_{e_j}}(i)$ for $E_j \in {\cal E}$, yielding
\begin{equation}\label{equa6}
{R_{e}}(i)
= \mathop {\max }\limits_{E_j \in {\cal{E}}} {R_{{e_j}}}(i) =  \mathop
{\max }\limits_{E_j \in {\cal{E}}} {\log _2}(1 +
\frac{{I|{h_{ie_j}}{|^2}}}{{|{h_{ip}}{|^2}{N_{e_j}}}}).
\end{equation}
As discussed in [13] and [14], the achievable secrecy rate is shown
as the difference between the capacities of the main link and the
wiretap link. Therefore, from Eqs. (4) and (6), the achievable
secrecy rate of cognitive transmission from $\textrm{CU}_i$ to CBS
in the presence of $N$ eavesdroppers is obtained as
\begin{equation}\label{equa7} {R_s}(i) =
\left[{R_b}(i) - \mathop {\max }\limits_{E_j \in {\cal{E}}}
{R_{{e_j}}}(i)\right]^{+},
\end{equation}
where subscript $s$ denotes `secrecy' and ${[x]^ + } = \max (x,0)$.
In general, a CU with the highest achievable secrecy rate should be
selected as the optimal user and scheduled for data transmission.
Hence, from Eq. (7), the user scheduling criterion can be given by
\begin{equation}\label{equa8}
\begin{split}
{\textrm{OptimalUser}}& = \arg \mathop {\max }\limits_{i \in {\cal{U}}} \left[ {{C_b}(i) - \mathop {\max }\limits_{j \in {\cal{E}}} {C_{{e_j}}}(i)} \right]\\
=\arg \mathop {\max }\limits_{i \in \cal{U}} &[ {{{\log }_2}(1 + \frac{{I|{h_{ib}}{|^2}}}{{|{h_{ip}}{|^2}{N_b}}}) } \\
& - {{\log }_2}(1 + \mathop {\max }\limits_{j \in {\cal{E}}} \frac{{I|{h_{i{e_j}}}{|^2}}}{{|{h_{ip}}{|^2}{N_{{e_j}}}}}) ],
\end{split}
\end{equation}
where $\cal{U}$ represents the set of $M$ CUs. One can observe from
Eq. (8) that not only the channel state information (CSI) of main
link $h_{ib}$, but also the wiretap link's CSI $h_{ie_j}$ is
considered in performing the multiuser scheduling. Thus, from Eq.
(8), the achievable secrecy rate of proposed multiuser scheduling
scheme with $M$ CUs in the presence of $N$ eavesdroppers is given by
\begin{equation}\label{equa9}
{R^{P}_s} = \mathop {\max }\limits_{i \in \cal{U}} {\left[ \begin{array}{l}
{{\log }_2}(1 +\dfrac{{I|{h_{ib}}{|^2}}}{{|{h_{ip}}{|^2}{N_b}}})  \\
  - {{\log }_2}(1 +\mathop {\max }\limits_{E_j \in {\cal{E}}}
\dfrac{{I|{h_{i{e_j}}}{|^2}}}{{|{h_{ip}}{|^2}{N_{{e_j}}}}}) \\
 \end{array} \right]^ + },
\end{equation}
where superscript $P$ denotes `proposed'. Notice that random
variables $|h_{ib}|^2$, $|h_{ip}|^2$ and $|h_{ie_j}|^2$ follow
exponential distributions with means $\sigma_{ib}^2$,
$\sigma_{ip}^2$ and $\sigma_{ie_j}^2$, respectively. For notational
convenience, $\sigma_{ib}^2$ and $\sigma_{ie_j}^2$ are,
respectively, denoted by $\sigma_{ib}^2=\theta_{ib}\sigma^2_{m}$ and
$\sigma_{ie_j}^2=\theta_{ie_j}\sigma^2_{e}$, where $\sigma^2_{m}$
and $\sigma^2_{e}$ represent the reference channel gains of the main
link and the wiretap link, respectively. Moreover, let
$\lambda_{me}=\sigma^2_{m}/\sigma^2_{e}$ denote the ratio of
$\sigma^2_{m}$ to $\sigma^2_{e}$, which is referred to as the
main-to-eavesdropper ratio (MER) throughout this paper. Denoting
$x_i=|h_{ib}|^2$, $y_i=|h_{ip}|^2$ and $z_{ij}=|h_{ie_j}|^2$, we can
easily obtain an ergodic secrecy rate of cognitive transmission from
Eq. (9) as
\begin{equation}\label{equa10}
\begin{split}
{\bar C^{P}_s} =&\underbrace {\iint {\cdots\int } }_{(M + 2)N}{{C_s^P\prod\limits_{i = 1}^M {(\frac{1}{{\sigma _{ib}^2\sigma _{ip}^2}}\prod\limits_{j = 1}^N {\frac{1}{{\sigma _{i{e_j}}^2}}} )}}}\\
&\quad \times \exp [ - \sum\limits_{i = 1}^M {(\frac{{{x_i}}}{{\sigma _{ib}^2}} + \frac{{{y_i}}}{{\sigma _{ip}^2}} + \sum\limits_{j = 1}^N {\frac{{{z_{ij}}}}{{\sigma _{i{e_j}}^2}}} )} ]d{x_i}d{y_i}d{z_{ij}},
\end{split}
\end{equation}
where $x_i>0$, $y_i>0$, and $z_{ij}>0$. It needs to be pointed out
that obtaining a closed-form solution to the high dimensional
integral in Eq. (10) is challenging, however the ergodic secrecy
rate of the proposed multiuser scheduling scheme can be numerically
determined through computer simulations.

\subsection{Traditional Multiuser Scheduling Scheme}
In this subsection, we present the traditional multiuser scheduling
scheme [19] as a benchmark scheme, where the main objective is to
maximize the achievable data rate at the desired destination CBS
without considering eavesdropping attacks. Thus, in the traditional
multiuser scheduling, a CU that maximizes the achievable rate at CBS
is viewed as the optimal user among $M$ CUs. Using Eq. (4), the
traditional multiuser scheduling criterion can be written as
\begin{equation}\label{equa11}
\begin{split}
\textrm{OptimalUser} &= \arg \mathop {\max }\limits_{U_i \in {\cal{U}}}R_{b}(i)\\
&=\arg \mathop {\max }\limits_{U_i \in {\cal{U}}} {\log _2}(1
+ \frac{{I|{h_{ib}}{|^2}}}{{|{h_{ip}}{|^2}{N_b}}}),
\end{split}
\end{equation}
from which the overall achievable rate at CBS using the traditional
multiuser scheduling scheme with $M$ CUs is given by
\begin{equation}\label{equa12}
R_b=\mathop {\max }\limits_{U_i \in {\cal{U}}} {\log _2}(
1 + \frac{{I|{h_{ib}}{|^2}}}{{|{h_{ip}}{|^2}{N_b}}}).
\end{equation}
For notational convenience, let `o' denote the optimal CU that is
selected by the traditional multiuser scheduling scheme. Similarly
to Eq. (6), the overall achievable rate at $N$ eavesdroppers from
the optimal CU is obtained as
\begin{equation}\label{equa13}
{R_{e}} = \mathop {\max }\limits_{E_j \in {\cal{E}}}
{\log _2}(1 + \frac{{I|{h_{oe_j}}{|^2}}}{{|{h_{op}}{|^2}{N_{e_j}}}}),
\end{equation}
where subscript `o' denotes the optimal CU. Combining Eqs. (12) and
(13), the achievable secrecy rate of the traditional multiuser
scheduling scheme is given by
\begin{equation}\label{equa14}
{R^{T}_s}= {\left[ \begin{array}{l}\mathop {\max }\limits_{U_i \in {\cal{U}}}{\log _2}
(1 + \dfrac{{I|{h_{ib}}{|^2}}}{{|{h_{ip}}{|^2}{N_b}}})\\
-\mathop {\max }
\limits_{E_j \in {\cal{E}}} {\log _2}(1 + \dfrac{{I|{h_{oe_j}}{|^2}}}
{{|{h_{op}}{|^2}{N_{e_j}}}})\\
 \end{array} \right]^ + },
\end{equation}
where superscript $T$ denotes `traditional'. Similarly to Eq. (10),
the ergodic secrecy rate of traditional multiuser scheduling scheme
can be obtained from Eq. (14) as
\begin{equation}\label{equa15}
\begin{split}
{\bar C^{T}_s} =&\underbrace {\iint {\cdots\int } }_{(M + 2)N}{{C_s^T\prod\limits_{i = 1}^M {(\frac{1}{{\sigma _{ib}^2\sigma _{ip}^2}}\prod\limits_{j = 1}^N {\frac{1}{{\sigma _{i{e_j}}^2}}} )}}}\\
&\quad \times \exp [ - \sum\limits_{i = 1}^M {(\frac{{{x_i}}}{{\sigma _{ib}^2}} + \frac{{{y_i}}}{{\sigma _{ip}^2}} + \sum\limits_{j = 1}^N {\frac{{{z_{ij}}}}{{\sigma _{i{e_j}}^2}}} )} ]d{x_i}d{y_i}d{z_{ij}},
\end{split}
\end{equation}
where $(x_i,y_i,z_{ij})>0$ and $R^{T}_s$ is given by Eq. (14).

\subsection{{Conventional Artificial Noise Scheme}}
This subsection presents the conventional artificial noise scheme
[21], [22] for the purpose of comparison with the user scheduling
approaches. The reasons for choosing the artificial noise scheme for comparison are twofold: (1) The artificial noise scheme is one of the most commonly used methods for the wireless physical-layer security,
which is often adopted as the basis of comparison; (2)
Although there are some different anti-eavesdropping techniques
(e.g., the artificial noise scheme [21], [22],
cooperative beamforming [27], resource allocation [28], etc.),
they each have their respective but complementary advantages.
Thus, for simplicity, we only consider the artificial noise approach
as the benchmark scheme. In the artificial noise scheme,
CUs are enabled to generate interfering signals
(called artificial noise) intelligently so that
only the eavesdroppers are adversely affected by
the interfering signals while the intended CBS is unaffected.
It has been shown in [21] that such artificial noise can be
designed to interfere with the eavesdroppers only without
affecting the legitimate receiver, if and only if the number
of antennas at the legitimate transmitter is more than the
number of antennas at the legitimate receiver. Since all
nodes as shown in Fig. 1 are equipped with single antenna,
we consider that $M$ CUs collaborate with each other and
share their antennas to form a virtual transmit antenna
array, which guarantees that the number of transmit antennas
is larger than the number of receive antennas at CBS for $M \ge 2$.
Without loss of generality, we denote the desired signal by $x$
which will be transmitted to CBS through the virtual transmit
antenna array. Meanwhile, the artificial noise vector is denoted
by ${\textrm{\emph{\textbf{w}}}}= \left( {w_1 ,w_2 , \cdots ,w_i ,
\cdots ,w_M } \right) $, where $w_i$ is to be transmitted by
$\textrm{CU}_i$. Notice that in the artificial noise scheme,
certain transmit power should be allocated to produce artificial
noise. The simple equal power allocation between the desired signal
and the artificial noise is shown as a near-optimal and effective
strategy [21], which is used throughout this paper. Hence,
considering that $M$ CUs simultaneously transmit the desired
signal $x$ and the artificial noise vector
${\textrm{\emph{\textbf{w}}}}$, the received signal at CBS can be expressed as
\begin{equation}\label{equa16}
y_b  = \sum\limits_{i = 1}^M {\sqrt {\frac{{P_i }}{2}} h_{ib} }
x + \sum\limits_{i = 1}^M {\sqrt {\frac{{P_i }}{2}} h_{ib} w_i }
+ n_b,
\end{equation}
where $P_i$ is the transmit power at $\textrm{CU}_i$ and $n_b$ is
the AWGN with zero mean and variance $N_b$. Since all CUs transmit
simultaneously, the total transmit power at CUs shall be constrained
to limit the interference received at PR. Given the maximum
tolerable interference $I$ at PR and $M$ CUs simultaneously
transmitting the desired signal and artificial noise, the
interference received at PU from each CU is limited by $ \frac{I}{M}
$ for equal allocation. Thus, the transmit power at $\textrm{CU}_i$,
$P_i$, is given by
\begin{equation}\label{equa17}
P_i  = \frac{I}{{M|h_{ip} |^2 }},
\end{equation}
where $h_{ip}$ represents the fading coefficient of the channel from
$\textrm{CU}_i$ to PU. Moreover, the artificial noise vector
${\textrm{\emph{\textbf{w}}}}$ should be designed to interfere with
the eavesdroppers only without affecting the intended CBS, implying
\begin{equation}\label{equa18}
\sum\limits_{i = 1}^M {\sqrt {\frac{{P_i }}{2}} h_{ib} w_i }  = 0,
\end{equation}
for ${\textrm{\emph{\textbf{w}}}} \ne 0$. The artificial noise
requirement as specified in Eq. (18) can be easily satisfied when
the number of CUs $M\ge2$. Substituting Eqs. (17) and (18) into Eq.
(16), we can obtain the achievable rate at CBS as
\begin{equation}\label{equa19}
R_b  = \log _2 (1 + \frac{I}{{2MN_b }}\left| {\sum\limits_{i = 1}
^M {\frac{{h_{ib} }}{{|h_{ip} |}}} } \right|^2 ).
\end{equation}
Meanwhile, the received signal at eavesdropper $E_j$ is written as
\begin{equation}\label{equa20}
y_{e_j}  = \sum\limits_{i = 1}^M {\sqrt {\frac{{P_i }}{2}} h_{ie_j} x}
+ \sum\limits_{i = 1}^M {\sqrt {\frac{{P_i }}{2}} h_{ie_j} w_i }
+ n_{e_j},
\end{equation}
where $n_{e_j}$ is a zero-mean AWGN noise of variance $N_{e_j}$
received at $E_j$. Since the wiretap channels $h_{ie_j}$ are
independent of the main channels $h_{ib}$, the artificial noise
satisfying Eq. (18) will result in harmful interference at the
eavesdropper $E_j$, i.e., $ \sum\limits_{i = 1}^M {\sqrt {\frac{{P_i
}}{2}} h_{ie_j} w_i } \ne 0$. Hence, the achievable rate at $E_j$
can be given by
\begin{equation}\label{equa21}
R_{e_j }  = \log _2 (1 + \frac{{I\left| {\sum\limits_{i = 1}^M
 {\frac{{h_{ie_j} }}{{|h_{ip} |}}} } \right|^2 }}{{I\left|
 {\sum\limits_{i = 1}^M {\frac{{h_{ie_j} }}{{|h_{ip} |}}} } \right|^2
  + 2MN_{e_j } }}),
\end{equation}
from which the overall achievable rate at $N$ eavesdroppers is given
by the highest one among $R_{e_j }$ for $1 \le j \le N$, yielding
\begin{equation}\label{equa22}
R_e  = \mathop {\max }\limits_{E_j \in {\cal E}} R_{e_j }  =
\mathop {\max }\limits_{E_j \in {\cal E}}\log _2 (1 +
\frac{{I\left| {\sum\limits_{i = 1}^M {\frac{{h_{ie_j} }}{{|h_{ip} |}}} }
 \right|^2 }}{{I\left| {\sum\limits_{i = 1}^M {\frac{{h_{ie_j} }}
 {{|h_{ip} |}}} } \right|^2  + 2MN_{e_j } }}).
\end{equation}
Combining Eqs. (19) and (22), the achievable secrecy rate of the
artificial noise scheme is given by
\begin{equation}\label{equa23}
\begin{split}
R_s^A  =  {\left[ \begin{array}{l}
 {\log _2 (1 + \dfrac{I}{{2MN_b }}\left| {\sum\limits_{i = 1}^M
{\dfrac{{h_{ib} }}{{|h_{ip} |}}} } \right|^2 )}  \\
  -{ \mathop {\max }\limits_{E_j \in {\cal E}} \log _2 (1 + \dfrac{{I\left| {\sum\limits_{i = 1}^M
 {\frac{{h_{ie_j} }}{{|h_{ip} |}}} } \right|^2 }}{{I\left|
 {\sum\limits_{i = 1}^M {\dfrac{{h_{ie_j} }}{{|h_{ip} |}}} }
 \right|^2  + 2MN_{e_j } }})} \\
 \end{array} \right]^ + }
\end{split}
\end{equation}
where superscript $A$ stands for `artificial noise'. A closed-form
expression of the ergodic secrecy rate for the artificial noise
scheme can be derived by averaging out the random variables
$h_{ib}$, $h_{ie_j}$ and $h_{ip}$ in Eq. (23), which is challenging
and cumbersome. Nevertheless, given the parameters $M$, $N$, $I$,
$N_b$, $N_{e_j}$, $\sigma^2_{ib}$, $\sigma^2_{ie_j}$ and
$\sigma^2_{ip}$, the ergodic secrecy rate may be readily determined
through computer simulations. So far, we have completed the
achievable secrecy rate analysis of the multiuser scheduling and the
artificial noise schemes. The following presents numerical secrecy
rate results to show the advantage of the proposed multiuser
scheduling over the traditional user scheduling and the artificial
noise schemes.

\subsection{Numerical Secrecy Rate Results}

\begin{figure}
  \centering
  {\includegraphics[scale=0.55]{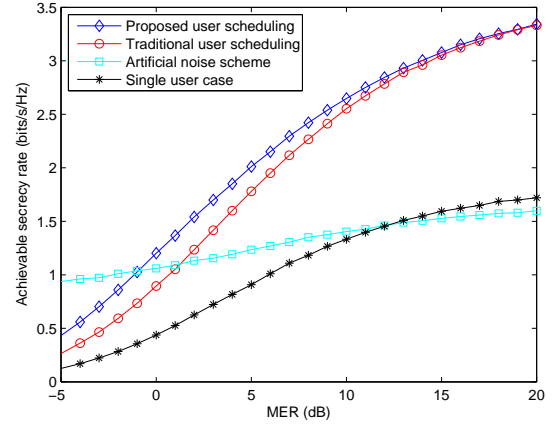}\\
  \caption{{Achievable secrecy rate versus MER of the single user transmission,
   the traditional multiuser scheduling, the artificial noise scheme,
    and the proposed multiuser scheduling with $M=4$, $N=2$, $I=N_{b}=N_{e_j}
     =0{\textrm{dBm}}$, $\sigma^2_{ib}=0.8$, $\sigma^2_{ip}=0.5$,
     $\theta_{ib}=1$, and $\theta_{i_{e_j}}=0.6$.}}\label{Fig2}}
\end{figure}
{This subsection presents the
numerical secrecy rate results of the traditional multiuser
scheduling, the artificial noise scheme and the proposed multiuser
scheduling. Fig. 2 shows the achievable secrecy rate comparison
among the single user transmission, the artificial noise scheme and
the multiuser scheduling approaches by using Eqs. (10), (15) and
(23). It is observed from Fig. 2 that in low MER region, the
artificial noise scheme performs better than both the traditional
and proposed multiuser scheduling approaches. As MER increases
beyond a critical value, the artificial noise scheme becomes worse
than the multiuser scheduling approaches, even worse than the single
user case in high MER region. This is because that with an
increasing MER, the wiretap link becomes much weaker than the main
link and the eavesdroppers will most likely fail to intercept the
legitimate transmissions. Therefore, as MER increases, the
eavesdroppers' channel conditions become worse and worse and thus it
is unnecessary to generate the artificial noise to confuse the
eavesdroppers in high MER region. However, the artificial noise
scheme wastes some power resources for producing the artificial
noise, which makes its achievable secrecy rate become lower than
that of the proposed multiuser scheduling scheme in high MER region.
In addition, one can see from Fig. 2 that the achievable secrecy
rate of the proposed multiuser scheduling is strictly higher than
that of the traditional multiuser scheduling across the whole MER
region, showing the advantage of the proposed multiuser scheduling
scheme. Moreover, as MER increases, the achievable secrecy rate
improvement of the proposed multiuser scheduling scheme over the
traditional multiuser scheduling becomes less notable. This is due
to the fact that for sufficiently large MERs, the wiretap link is
negligible as compared with the main link and thus the achievable
secrecy rate at CBS through the main link dominates the achievable
secrecy rates of Eqs. (9) and (14), leading to the convergence of
the achievable secrecy rates between the proposed and traditional
multiuser scheduling schemes.}

\begin{figure}
  \centering
  {\includegraphics[scale=0.55]{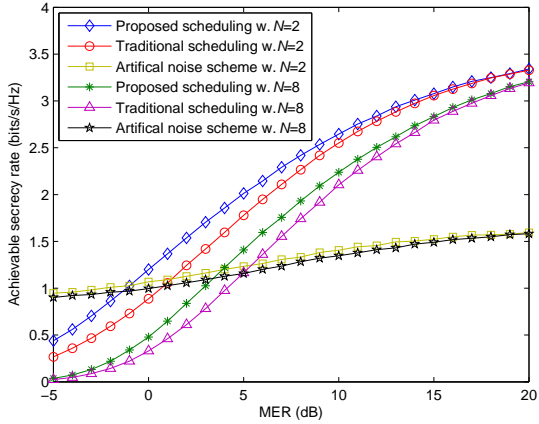}\\
  \caption{{Achievable secrecy rate versus MER of the traditional multiuser scheduling, the artificial noise scheme,
    and the proposed multiuser scheduling for different number of eavesdroppers $N$ with $M=4$,
   $I=N_{b}=N_{e_j} =0{\textrm{dBm}}$,
   $\sigma^2_{ib}=0.8$, $\sigma^2_{ip}=0.5$,
     $\theta_{ib}=1$, and $\theta_{i_{e_j}}=0.6$.}}\label{Fig3}}
\end{figure}
{In Fig. 3, we show the achievable
secrecy rate versus MER of the traditional and proposed multiuser
scheduling schemes as well as the artificial noise approach for
different number of eavesdroppers $N$ with $M=4$ and
$I=N_{b}=N_{e_j} =0{\textrm{dBm}}$. As shown in Fig. 3, for both
cases of $N=2$ and $N=8$, the traditional and proposed multiuser
scheduling schemes initially have lower secrecy rate than the
artificial noise scheme in low MER region. As MER continues
increasing beyond a certain value, the traditional and proposed
multiuser scheduling schemes finally outperform the artificial noise
scheme in terms of the achievable secrecy rate. Fig. 3 also
demonstrates that as the number of eavesdroppers increases from
$N=2$ to $N=8$, the achievable secrecy rates of the traditional and
proposed multiuser scheduling schemes are significantly reduced. In
contrast, the achievable secrecy rate of the artificial noise scheme
decreases non-significantly. This is because that the artificial
noise scheme generates significant interferences against
eavesdropping attacks, which makes its achievable secrecy rate
robust to the eavesdroppers' channel conditions. In addition, one
can see from Fig. 3 that the proposed multiuser scheduling scheme
always performs better than the traditional multiuser scheduling in
terms of the achievable secrecy rate, which further confirms the
advantage of the proposed multiuser scheduling scheme.}
\begin{figure}
  \centering
  {\includegraphics[scale=0.55]{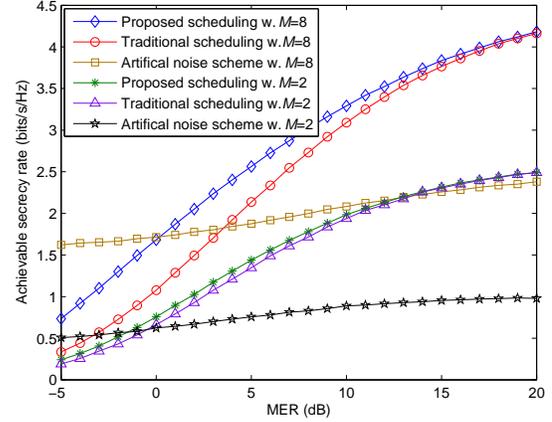}\\
  \caption{{Achievable secrecy rate versus MER of the traditional
  multiuser scheduling, the artificial noise scheme,
    and the proposed multiuser scheduling for different
  number of CUs $M$ with $N=2$, $I=N_{b}=N_{e_j} =0{\textrm{dBm}}$,
    $\sigma^2_{ib}=0.8$, $\sigma^2_{ip}=0.5$,
     $\theta_{ib}=1$, and $\theta_{i_{e_j}}=0.6$.}}\label{Fig4}}
\end{figure}

{Fig. 4 shows the achievable secrecy
rate versus MER of the traditional and proposed multiuser scheduling
schemes as well as the artificial noise approach for different
number of CUs $M$ with $N=2$ and $I=N_{b}=N_{e_j} =0{\textrm{dBm}}$.
One can observe from Fig. 4 that for both cases of $M=2$ and $M=8$,
the achievable secrecy rates of the traditional and proposed
multiuser scheduling schemes are lower than that of the artificial
noise scheme in low MER region. However, when MER is larger than a
certain value and continues increasing, the traditional and proposed
multiuser scheduling schemes significantly outperform the artificial
noise scheme in terms of the achievable secrecy rate. It can also be
seen from Fig. 4 that as the number of CUs increases from $M=2$ to
$M=8$, the achievable secrecy rates of both the traditional and
proposed multiuser scheduling schemes increase significantly.
Although the secrecy rate of the traditional multiuser scheduling is
lower than that of the proposed multiuser scheduling, its secrecy
rate also improves as the number of CUs increases from $M=2$ to
$M=8$, showing the security enhancement of exploiting multiuser
scheduling even when the eavesdroppers' CSI is unavailable.
Therefore, if the eavesdroppers' CSI is unknown, we can consider the
use of traditional multiuser scheduling scheme that does not require
the eavesdroppers' CSI. If the eavesdroppers' CSI becomes available,
the proposed multiuser scheduling would be a better choice.}

\section{Intercept Probability Analysis over Rayleigh Fading Channels}
In this section, we analyze the intercept probability of the
traditional and proposed multiuser scheduling schemes as well as the
artificial noise scheme over Rayleigh fading channels. We also
provide numerical results on the intercept probability to show the
advantage of proposed multiuser scheduling scheme over the
traditional multiuser scheduling and the artificial noise schemes.
\subsection{Proposed Multiuser Scheduling Scheme}
This subsection presents the intercept probability analysis of
proposed multiuser scheduling scheme. As discussed in [12] and [13],
an intercept event occurs when the achievable secrecy rate of the
main link becomes less than that of the wiretap link. Thus, from Eq.
(9), we can obtain an intercept probability of the cognitive
transmission with the proposed multiuser scheduling scheme as Eq.
(24) at the top of the following page.
\begin{figure*}
\begin{equation}\label{equal24}
P_{{\mathop{\rm int}} }^P = \Pr \left\{ {\mathop {\max }
\limits_{U_i \in \cal{U}} \left[ {{{\log }_2}(1
+ \frac{{I|{h_{ib}}{|^2}}}{{|{h_{ip}}{|^2}{N_b}}})
 - {{\log }_2}(1 + \mathop {\max }\limits_{E_j \in
 {\cal{E}}} \frac{{I|{h_{i{e_j}}}{|^2}}}{{|{h_{ip}}
 {|^2}{N_{{e_j}}}}})} \right] < 0} \right\}.
\end{equation}
\end{figure*}
Notice that random variables $|h_{ib}|^2$, $|h_{ip}|^2$ and
$|h_{ie_j}|^2$ are independent of each other for different
$\textrm{CU}_i$. Thus, the intercept probability of proposed
multiuser scheduling scheme can be computed from Eq. (24) as
\begin{equation}\label{equa25}
P_{{\mathop{\rm int}} }^P= \prod\limits_{i = 1}^M {\Pr \left( {\frac{{|{h_{ib}}{|^2}}}
 {{{N_b}}} < \mathop {\max }\limits_{E_j \in {\cal{E}}}
  \frac{{|{h_{i{e_j}}}{|^2}}}{{{N_{{e_j}}}}}} \right)},
\end{equation}
where $M$ is the number of CUs. Considering that $|h_{ib}|^2$ and
$|h_{ie_j}|^2$ are independent exponentially distributed random
variables with respective means $\sigma_{ib}^2$ and
$\sigma_{ie_j}^2$ and letting $x=|h_{ib}|^2$, we can obtain
\begin{equation}\label{equa26}
\begin{split}
&\Pr \left( {\frac{{|{h_{ib}}{|^2}}}{{{N_b}}} <
 \mathop {\max }\limits_{E_j \in {\cal{E}}} \frac{
 {|{h_{i{e_j}}}{|^2}}}{{{N_{{e_j}}}}}} \right) \\
 &=1 - \Pr \left( {\mathop {\max }\limits_{E_j \in
 {\cal{E}}} \frac{{|{h_{i{e_j}}}{|^2}}}{{{N_{{e_j}}}}}
 < \frac{x}{{{N_b}}}} \right)\\
&=1- \int^{\infty}_{{ 0}}{{\prod\limits_{j = 1}^N
{[1 - \exp ( - \frac{{{N_{{e_j}}}x}}{{{N_b}\sigma _
{i{e_j}}^2}})]} \frac{1}{{\sigma _{ib}^2}}\exp
( - \frac{x}{{\sigma _{ib}^2}})dx}},
\end{split}
\end{equation}
where $N$ is the number of eavesdroppers. Using the binomial
theorem, we can expand term $\prod\limits_{j = 1}^N {[1 - \exp ( -
\frac{{{N_{{e_j}}}x}}{{{N_b}\sigma _{i{e_j}}^2}})]} $ as
\begin{equation}\label{equa27}
\begin{split}
&\prod\limits_{j = 1}^N {[1 - \exp ( - \frac{{{N_{{e_j}}}x}}
{{{N_b}\sigma _{i{e_j}}^2}})]}  \\
&= 1 + \sum\limits_
{n = 1}^{{2^N} - 1} {{{( - 1)}^{|{{\cal E}_n}|}}\exp ( -
\sum\limits_{E_j \in {{\cal E}_n}} {\frac{{{N_{{e_j}}}x}}{{{N_b}\sigma _{i{e_j}}^2}}} )},
\end{split}
\end{equation}
where ${\cal E}_n$ is the $n{\textrm{-th}}$ non-empty subcollection
of $N$ eavesdroppers and $|{\cal E}_n|$ represents the cardinality
of set ${\cal E}_n$. Substituting Eq. (27) into Eq. (26) and
performing the integration yield
\begin{equation}\label{equa28}
\begin{split}
&\Pr \left( {\frac{{|{h_{ib}}{|^2}}}{{{N_b}}} <
\mathop {\max }\limits_{E_j \in {\cal{E}}} \frac{{
|{h_{i{e_j}}}{|^2}}}{{{N_{{e_j}}}}}} \right) \\
&=\sum\limits_{n = 1}^{{2^N} - 1} {{{( - 1)}^{|{{\cal E}_n}|
+ 1}}{(1 + \sum\limits_{E_j \in {{\cal E}_n}} {\frac{{{N_{{e_j}}}
\sigma _{ib}^2}}{{{N_b}\sigma _{i{e_j}}^2}}} )^{ - 1}}}.
\end{split}
\end{equation}
Combining Eqs. (25) and (28), we obtain a closed-form expression of
the intercept probability for the proposed multiuser scheduling
scheme as
\begin{equation}\label{equa29}
P_{{\mathop{\rm int}} }^P = \prod\limits_{i = 1}^M
 {\left[ {\sum\limits_{n = 1}^{{2^N} - 1} {{{( - 1)}
 ^{|{{\cal E}_n}| + 1}}{(1 + \sum\limits_{E_j \in {{\cal E}_n}}
 {\frac{{{N_{{e_j}}}\sigma _{ib}^2}}{{{N_b}\sigma _{i{e_j}}^2}}} )^{ - 1}}} } \right]}.
\end{equation}
Denoting $\sigma_{ib}^2=\theta_{ib}\sigma^2_{m}$,
$\sigma_{ie_j}^2=\theta_{ie_j}\sigma^2_{e}$, and
$\lambda_{me}=\sigma^2_{m}/\sigma^2_{e}$, the preceding equation can
be rewritten as
\begin{equation}\label{equa30}
P_{{\mathop{\rm int}} }^P = \prod\limits_{i = 1}^M
{\left[ {\sum\limits_{n = 1}^{{2^N} - 1} {{{( - 1)}^
{|{{\cal E}_n}| + 1}}{(1 + \sum\limits_{E_j \in {{\cal E}_n}} {\frac
{{{N_{{e_j}}}{\theta _{ib}}}}{{{N_b}{\theta _{i{e_j}}}}}
{\lambda _{me}}} )^{ - 1}}} } \right]},
\end{equation}
where $\lambda_{me}=\sigma^2_{m}/\sigma^2_{e}$ is called
main-to-eavesdropper ratio (MER) throughout this paper.

\subsection{Traditional Multiuser Scheduling Scheme}
In this subsection, we analyze the intercept probability of
traditional multiuser scheduling scheme for the comparison purpose.
From Eqs. (12) and (13), an intercept probability of the cognitive
transmission relying on the traditional multiuser scheduling scheme
is given by
\begin{equation}\label{equa31}
\begin{array}{l}
 P_{{\mathop{\rm int}} }^T = \Pr ({R_b} < {R_e}) \\
  \quad\quad= \Pr \left[ \begin{array}{l}
 \mathop {\max }\limits_{{U_i} \in \cal{U}} {\log _2}(1 + \dfrac{{I|{h_{ib}}{|^2}}}{{|{h_{ip}}{|^2}{N_b}}}) \\
  < \mathop {\max }\limits_{{E_j} \in \cal{E}} {\log _2}(1 + \dfrac{{I|{h_{o{e_j}}}{|^2}}}{{|{h_{op}}{|^2}{N_{{e_j}}}}}) \\
 \end{array} \right] \\
 \end{array},
\end{equation}
which can be further simplified to
\begin{equation}\label{equa32}
P_{{\mathop{\rm int}} }^T = \Pr \left[ {\mathop
{\max }\limits_{U_i \in {\cal{U}}} \frac{{|{h_{ib}}
{|^2}}}{{|{h_{ip}}{|^2}{N_b}}} < \mathop {\max }
\limits_{E_j \in {\cal{E}}} \frac{{|{h_{o{e_j}}}{|^2}}}
{{|{h_{op}}{|^2}{N_{{e_j}}}}}} \right].
\end{equation}
Although obtaining a general closed-form solution to Eq. (32) for
any $M$ and $N$ is difficult, numerical intercept probabilities of
the traditional multiuser scheduling scheme can be easily determined
through computer simulations. For illustration purposes, the
following presents the intercept probability analysis for a special
case with the single CU (i.e., $M=1$). Substituting $M=1$ into Eq.
(32) gives
\begin{equation}\label{equa33}
\begin{split}
P_{{\mathop{\rm int}} }^T &= \Pr \left[ {\frac{{|{h_{1b}}{|^2}}}
{{|{h_{1p}}{|^2}{N_b}}} < \mathop {\max }\limits_{E_j \in {\cal{E}}}
\frac{{|{h_{1{e_j}}}{|^2}}}{{|{h_{1p}}{|^2}{N_{{e_j}}}}}} \right]\\
&= 1-\Pr \left[ {\frac{{|{h_{1b}}{|^2}}}{{{N_b}}} > \mathop
{\max }\limits_{E_j \in {\cal{E}}} \frac{{|{h_{1{e_j}}}{|^2}}}{{{N_{{e_j}}}}}} \right].
\end{split}
\end{equation}
Notice that random variables $|h_{1b}|^2$ and $|h_{1e_j}|^2$ follow
exponential distributions with respective means $\sigma_{1b}^2$ and
$\sigma_{1e_j}^2$ and are independent of each other. Denoting
$x=|h_{1b}|^2$, we can obtain
\begin{equation}\label{equa34}
\begin{split}
P_{{\mathop{\rm int}} }^T &= 1 - \Pr \left[ {\frac{x}{{{N_b}}}
 > \mathop {\max }\limits_{E_j \in {\cal{E}}}
 \frac{{|{h_{1{e_j}}}{|^2}}}{{{N_{{e_j}}}}}} \right]\\
& = 1 - \int_0^\infty  {\prod\limits_{j = 1}^N
 {[1 - \exp ( - \frac{{{N_{{e_j}}}x}}{{\sigma _{1{e_j}}
 ^2{N_b}}})]} \frac{1}{{\sigma _{1b}^2}}
 \exp ( - \frac{x}{{\sigma _{1b}^2}})dx}\\
& = \sum\limits_{n = 1}^{{2^N} - 1} {{{( - 1)}
^{|{{\cal E}_n}| + 1}}{(1 + \sum\limits_{E_j \in {{\cal E}_n}}
{\frac{{{N_{{e_j}}}\sigma _{1b}^2}}{{{N_b}\sigma _{1{e_j}}^2}}} )^{ - 1}}} ,
\end{split}
\end{equation}
for $M=1$, where the last equation is obtained by using the binomial
expansion formula, ${\cal E}_n$ is the $n{\textrm{-th}}$ non-empty
subcollection of $N$ eavesdroppers, and $|{\cal E}_n|$ represents
the cardinality of set ${\cal E}_n$.

\subsection{{Conventional Artificial Noise Scheme}}
{This subsection presents the
intercept probability analysis of the artificial noise scheme. Using
Eq. (23), we obtain the intercept probability of the artificial
noise scheme as
\begin{equation}\label{equa35}
P_{{\mathop{\rm int}} }^A  = \Pr \left[ \begin{array}{l}
 \dfrac{I}{{2MN_b }}\left|{\sum\limits_{i = 1}^M {\dfrac{{h_{ib} }}{{|h_{ip} |}}} } \right|^2  \\
  < \mathop {\max }\limits_{E_j \in {\cal E}} \dfrac{{I\left| {\sum\limits_{i =1}^M {\dfrac{{h_{ie_j } }}{{|h_{ip} |}}} } \right|^2 }}{{I\left|{\sum\limits_{i = 1}^M {\dfrac{{h_{ie_j } }}{{|h_{ip} |}}} }\right|^2  + 2MN_{e_j } }}  \\
 \end{array} \right],
\end{equation}
which can be used to compute the numerical intercept probability of
the artificial noise scheme. Moreover, using inequality $ {I\left|
{\sum\limits_{i = 1}^M {\frac{{h_{ie_j } }}{{|h_{ip} |}}} }
\right|^2  + 2MN_{e_j } } >  {2MN_{e_j } }$, we obtain an upper
bound on the intercept probability $P_{{\mathop{\rm int}} }^A$ as
\begin{equation}\label{equa36}
\begin{split}
 P_{{\mathop{\rm int}} }^A & < \Pr \left[ {\frac{I}{{2MN_b }}\left|
 {\sum\limits_{i = 1}^M {\frac{{h_{ib} }}{{|h_{ip} |}}} }
 \right|^2  < \mathop {\max }\limits_{E_j \in {\cal E}} \frac{I}
 {{2MN_{e_j } }}\left| {\sum\limits_{i = 1}^M {\frac{
 {h_{ie_j } }}{{|h_{ip} |}}} } \right|^2 } \right] \\
 & = \Pr \left[ {\frac{1}{{N_b }}\left| {\sum\limits_{i = 1}^M
 {\frac{{h_{ib} }}{{|h_{ip} |}}} } \right|^2  < \mathop
 {\max }\limits_{E_j \in {\cal E}} \frac{1}{{N_{e_j } }}\left|
 {\sum\limits_{i = 1}^M {\frac{{h_{ie_j } }}
 {{|h_{ip} |}}} } \right|^2 } \right].
 \end{split}
\end{equation}
Considering a special case of $M=1$, we can simplify Eq. (36) as
\begin{equation}\label{equa37}
P_{{\mathop{\rm int}} }^A  < \Pr \left[ {\frac{{|h_{1b} |^2 }}
{{|h_{1p} |^2 N_b }} < \mathop {\max }\limits_{E_j \in {\cal E}}
 \frac{{|h_{1e_j } |^2 }}{{|h_{1p} |^2 N_{e_j } }}} \right].
\end{equation}
Substituting $ P_{{\mathop{\rm int}} }^T  = \Pr \left[
{\frac{{|h_{1b} |^2 }}{{|h_{1p} |^2 N_b }} < \mathop {\max
}\limits_{E_j \in {\cal E}} \frac{{|h_{1e_j } |^2 }}{{|h_{1p} |^2
N_{e_j } }}} \right] $ from Eq. (33) into Eq. (37) yields
\begin{equation}\label{equa38}
P_{{\mathop{\rm int}} }^A  < P_{{\mathop{\rm int}} }^T,
\end{equation}
for $M=1$. This theoretically proves the intercept probability of
the artificial noise scheme is strictly lower than that of the
traditional user scheduling scheme for $M=1$. In what follows, we
show the numerical intercept probabilities of the traditional and
proposed user scheduling schemes as well as the artificial noise
approach.}

\subsection{Numerical Intercept Probability Results}
\begin{figure}
  \centering
  {\includegraphics[scale=0.55]{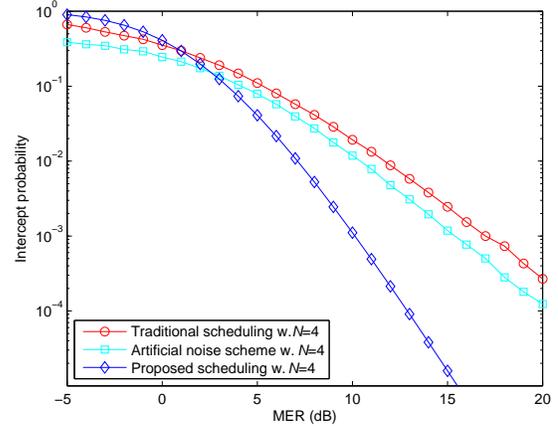}\\
  \caption{{Intercept probability versus MER of the traditional and proposed multiuser
   scheduling schemes as well as the artificial noise approach with $M=N=4$, $I=N_{b}=N_{e_j} =0{\textrm{dBm}}$,
    $\sigma^2_{ib}=\sigma^2_{ip}=1$, and $\theta_{ib}=\theta_{i_{e_j}}=1$.}}\label{Fig5}}
\end{figure}
{This subsection presents the
numerical results on intercept probability of the traditional and
proposed multiuser scheduling schemes as well as the artificial
noise scheme. In Fig. 5, we show the intercept probability versus
MER of the multiuser scheduling and the artificial noise schemes
with $M=N=4$ and $I=N_{b}=N_{e_j} =0{\textrm{dBm}}$. It is seen from
Fig. 5 that the intercept probability of the proposed multiuser
scheduling scheme is smaller than that of the traditional multiuser
scheduling and the artificial noise schemes, showing the advantage
of the proposed multiuser scheduling over the conventional
approaches. Fig. 5 also shows that the artificial noise scheme has
lower intercept probability than the traditional multiuser
scheduling scheme, which confirms the result of Eq. (38). In
addition, Fig. 5 demonstrates that that the slope of intercept
probability curve of the proposed multiuser scheduling scheme in
high MER region is much steeper that that of the traditional
multiuser scheduling and the artificial noise schemes. This means
that with an increasing MER, the intercept probability of the
proposed multiuser scheduling scheme is reduced at much higher speed
than that of the traditional multiuser scheduling and the artificial
noise schemes.}

\begin{figure}
  \centering
  {\includegraphics[scale=0.55]{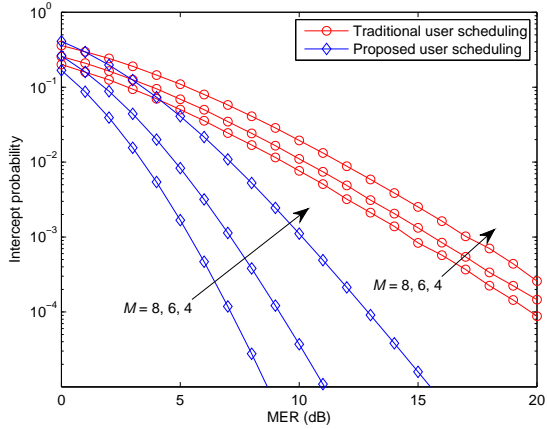}\\
  \caption{Intercept probability versus MER of the traditional and proposed multiuser
  scheduling schemes for different number of CUs $M$ with
   $N=4$, $I=N_{b}=N_{e_j} =0{\textrm{dBm}}$, $\sigma^2_{ib}
   =\sigma^2_{ip}=1$, and $\theta_{ib}=\theta_{i_{e_j}}=1$.}\label{Fig6}}
\end{figure}

Fig. 6 shows the intercept probability versus MER of the traditional
and proposed multiuser scheduling schemes for different number of
CUs $M$ with $N=4$, where $M=4$, $M=6$ and $M=8$ are considered for
illustration. As shown in Fig. 6, for all cases of $M=4$, $M=6$ and
$M=8$, the proposed multiuser scheduling scheme significantly
outperforms the traditional scheme in terms of intercept
probability, especially in high MER region. Also, as the number of
CUs increases from $M=4$ to $M=8$, the intercept probabilities of
the proposed and traditional multiuser scheduling schemes are
reduced significantly. Therefore, increasing the number of CUs can
effectively improve the security performance of cognitive
transmission. This confirms the security benefits by exploiting the
multiuser scheduling to defend against eavesdropping attacks in
cognitive radio networks.

\section{Diversity Order Analysis}
In this section, we analyze the diversity order of proposed
multiuser scheduling scheme to provide an insight into the impact of
the number of CUs on the intercept probability performance of
cognitive transmission in the presence of $N$ eavesdroppers. First,
let us recall the definition of traditional diversity order used to
evaluate the wireless transmission reliability performance in [29],
which is given by
\begin{equation}\label{equa39}
d =  - \mathop {\lim }\limits_{{\textrm{SNR}} \to \infty } \frac{\log {P_e}({\textrm{SNR}})}{\log {\textrm{SNR}}},
\end{equation}
where $\textrm{SNR}$ stands for signal-to-noise ratio and
$P_e({\textrm{SNR}})$ represents bit error rate (BER). In the
geometric sense, the traditional diversity order is to characterize
the slope of BER curve as SNR tends to infinity. However, in the
presence of eavesdropping attacks, the intercept probability is used
to evaluate the wireless security of cognitive transmission.
Moreover, it is observed from Eq. (30) that the intercept
probability is independent of signal power, which makes the
traditional diversity order definition become inapplicable to the
cognitive transmission scenario with multiple eavesdroppers.
Considering the fact that MER is a dominant factor in determining
the intercept probability of cognitive transmission, we define a
so-called security diversity as an asymptotic ratio of the
logarithmic intercept probability to logarithmic MER $\lambda_{me}$
with $\lambda_{me} \to \infty $. Accordingly, the diversity order of
the proposed multiuser scheduling scheme is given by
\begin{equation}\label{equa40}
d = -\mathop {\lim }\limits_{{\lambda _{me}} \to \infty }
\dfrac{{\log (P_{{\rm{int}}}^{P})}}{{\log ({\lambda _{me}})}},
\end{equation}
where $P_{{\rm{int}}}^{P}$ represents the intercept probability of
the proposed multiuser scheduling scheme. Letting $X = \mathop {\max
}\limits_{E_j \in {\cal{E}}}
\frac{{|{h_{i{e_j}}}{|^2}}}{{{N_{{e_j}}}}}$ wherein random variables
$|h_{ie_j}|^2$ ($j=1,2,\cdots,N$) follow independent exponential
distributions with respective means $\sigma^2_{ie_j}$, we obtain the
cumulative distribution function (CDF) of $X$ as
\begin{equation}\label{equa41}
\begin{split}
{P_X}(x) &= \Pr \left( {X < x} \right) \\
&= \Pr \left( {\mathop {\max }\limits_{E_j \in {\cal{E}}} \frac{{|{h_{i{e_j}}}{|^2}}}{{{N_{{e_j}}}}} < x} \right) \\
&= \prod\limits_{j = 1}^N {\left[ {1 - \exp ( - \frac{{{N_{{e_j}}}x}}{{\sigma _{i{e_j}}^2}})} \right]} ,
\end{split}
\end{equation}
which can be rewritten as
\begin{equation}\label{equa42}
{P_X}(x) = 1 + \sum\limits_{n = 1}^{{2^N} - 1} {{{( - 1)}^{|{{\cal E}_n}|}}\exp ( - \sum\limits_{E_j \in {{\cal E}_n}} {\frac{{{N_{{e_j}}}x}}{{\sigma _{i{e_j}}^2}}} )},
\end{equation}
where ${\cal E}_n$ is the $n{\textrm{-th}}$ non-empty subcollection
of $N$ eavesdroppers and $|{\cal E}_n|$ represents the cardinality
of set ${\cal E}_n$. From Eq. (42), the probability density function
(PDF) of $X$ is given by
\begin{equation}\label{equa43}
{p_X}(x) = \sum\limits_{n = 1}^{{2^N} - 1} {{{( - 1)}^{|{{\cal E}_n}| + 1}}\sum\limits_{E_j \in {{\cal E}_n}} {\frac{{{N_{{e_j}}}}}{{\sigma _{i{e_j}}^2}}} \exp ( - \sum\limits_{E_j \in {{\cal E}_n}} {\frac{{{N_{{e_j}}}x}}{{\sigma _{i{e_j}}^2}}} )}.
\end{equation}
Hence, using Eqs. (25) and (43), we can obtain $P_{{\rm{int}}}^{P}$
as
\begin{equation}\label{equa44}
\begin{split}
P_{{\mathop{\rm int}} }^P &= \prod\limits_{i = 1}^M {\Pr \left( {\frac{{|{h_{ib}}{|^2}}}{{{N_b}}} < X} \right)}\\
&= \prod\limits_{i = 1}^M {\int_0^\infty  {[1 - \exp ( - \frac{{{N_b}x}}{{\sigma _{ib}^2}})]{p_X}(x)dx} }.
\end{split}
\end{equation}
Letting $\lambda_{me} \to \infty$ and using Appendix A, we have
\begin{equation}\label{equa45}
1 - \exp ( - \frac{{{N_b}x}}{{\sigma _{ib}^2}}) = \frac{{{N_b}x}}{{\sigma _{ib}^2}} + O(\frac{1}{{{\lambda _{me}}}}),
\end{equation}
where $O(\frac{1}{{{\lambda _{me}}}})$ represents high-order
infinitesimals. Hence, substituting Eqs. (43) and (45) into Eq. (44)
and ignoring the high-order terms yield
\begin{equation}\label{equa46}
\begin{split}
P_{{\mathop{\rm int}} }^P &= \prod\limits_{i = 1}^M {\left[ {\sum\limits_{n = 1}^{{2^N} - 1} {{{( - 1)}^{|{{\cal E}_n}| + 1}}\int_0^\infty  {\frac{{{N_b}}}{{\sigma _{ib}^2}}\exp ( - \sum\limits_{E_j \in {{\cal E}_n}} {\frac{{{N_{{e_j}}}x}}{{\sigma _{i{e_j}}^2}}} )dx} } } \right]} \\
&= \prod\limits_{i = 1}^M {\left[ {\sum\limits_{n = 1}^{{2^N} - 1} {{{( - 1)}^{|{{\cal E}_n}| + 1}}{(\sum\limits_{E_j \in {{\cal E}_n}} {\frac{{{N_{{e_j}}}\sigma _{ib}^2}}{{{N_b}\sigma _{i{e_j}}^2}}} )^{ - 1}}} } \right]},
\end{split}
\end{equation}
for $\lambda_{me} \to \infty$. Denoting
$\sigma^2_{ib}=\theta_{ib}\sigma^2_{m}$ and
$\sigma^2_{ie_j}=\theta_{ie_j}\sigma^2_{e}$ and letting
$\lambda_{me} \to \infty$, we can rewrite $P_{{\mathop{\rm int}} }^P
$ from Eq. (46) as
\begin{equation}\label{equa47}
P_{{\mathop{\rm int}} }^P = \prod\limits_{i = 1}^M {\left[ {\sum\limits_{n = 1}^{{2^N} - 1} {{{( - 1)}^{|{{\cal E}_n}| + 1}}{(\sum\limits_{E_j \in {{\cal E}_n}} {\frac{{{N_{{e_j}}}{\theta _{ib}}}}{{{N_b}{\theta _{i{e_j}}}}}} )^{ - 1}}} } \right]}  \cdot {\left( {\frac{1}{{{\lambda _{me}}}}} \right)^M}.
\end{equation}
One can observe from Eq. (47) that the intercept probability of the
proposed multiuser scheduling scheme behaves as
${(\frac{1}{{{\lambda _{me}}}})^M}$ for $\lambda_{me} \to \infty$.
Substituting Eq. (47) into Eq. (40) yields
\begin{equation}\label{equa36}
d = M,
\end{equation}
which shows that the diversity order is the same as the number of
CUs, implying the full diversity achieved by the proposed multiuser
scheduling scheme. One can also see from Eq. (48) that the diversity
order is independent of the number of eavesdroppers, i.e., the
security diversity of proposed multiuser scheduling scheme is
insusceptible to the number of eavesdroppers. To be specific,
although increasing the number of eavesdroppers would definitely
degrade the intercept probability performance, it won't affect the
speed at which the intercept probability decreases as $\lambda_{me}
\to \infty$. In contrast, as the number of CUs increases, the slope
of intercept probability curve of the proposed multiuser scheduling
becomes steeper as $\lambda_{me} \to \infty$ in the geometric sense.
In other words, as $\lambda_{me} \to \infty$, the intercept
probability of proposed multiuser scheduling scheme decreases at
faster speed with an increasing number of CUs. Therefore, exploiting
multiuser scheduling can effectively improve the physical-layer
security of cognitive transmission to defend against eavesdropping
attacks.

\section{Conclusion}
In this paper, we have explored the physical-layer security of
cognitive transmissions in the presence of multiple eavesdroppers
and proposed the multiuser scheduling scheme to improve the
cognitive transmission security against eavesdropping attacks. For
the comparison purpose, we have studied the traditional multiuser
scheduling and the artificial noise approach as benchmark schemes.
We have analyzed the achievable secrecy rates of the traditional and
proposed multiuser scheduling schemes as well as the artificial
noise scheme with a maximum tolerable interference constraint for
the primary QoS protection. It has been shown that given a primary
QoS constraint, the achievable secrecy rates of the traditional and
proposed multiuser scheduling schemes are smaller than that of the
artificial noise scheme in low MER region. As MER increases beyond a
critical value, both the traditional and proposed multiuser
scheduling schemes have higher achievable secrecy rates than the
artificial noise scheme. Moreover, the proposed multiuser scheduling
scheme always outperforms the traditional multiuser scheduling in
terms of the achievable secrecy rate. We have also derived the
closed-form intercept probability expressions of the multiuser
scheduling and the artificial noise schemes and demonstrated that
the intercept performance of the proposed multiuser scheduling
scheme is better than that of the traditional multiuser scheduling
and the artificial noise schemes. In addition, we have examined the
diversity order performance through an asymptotic intercept
probability analysis and shown that the full diversity is achieved
by the proposed multiuser scheduling scheme, showing the advantage
of exploiting multiuser scheduling for enhancing the cognitive
transmission security against eavesdropping attacks.

It is worth mentioning that in this paper, we have not considered user fairness in the multiuser scheduling for improving the cognitive radio security against eavesdropping attacks. For example, if a cognitive user experiences severe propagation loss in a shadow fading environment, it will be rarely scheduled for accessing the channel by using the proposed multiuser scheduling scheme, causing a long channel access delay. It is thus of high practical interest to extend the results of this paper e.g. using the QoS guaranteed user scheduling (e.g., the proportional fair scheduling [30]). More specifically, user fairness should be further considered in the QoS guaranteed scheduling, attempting to maximize the achievable secrecy rate while at the same time guaranteeing each user with certain opportunities to access the channel. We will leave the above interesting problem for our future work.

\appendices
\section{Proof of Eq. (45)}
For notational convenience, we denote $t = \frac{{{N_b}x}}{{\sigma _{ib}^2}}$ where the PDF of random variable $x$ is given by Eq. (31). The mean of $t$ is expressed as
\begin{equation}
E(t) = \int_0^\infty  {\frac{{{N_b}x}}{{\sigma _{ib}^2}}{p_X}(x)dx}.\tag{A.1}\label{A.1}
\end{equation}
Substituting Eq. (43) into Eq. (A.1) yields
\begin{equation}
\begin{split}
E(t)= \sum\limits_{n = 1}^{{2^N} - 1} {{{( - 1)}^{|{E_n}| + 1}}{(\sum\limits_{j \in {E_n}} {\frac{{{N_{{e_j}}}\sigma _{ib}^2}}{{{N_b}\sigma _{i{e_j}}^2}}} )^{ - 1}}}.
\end{split}\tag{A.2}\label{A.2}
\end{equation}
Denoting $\sigma^2_{ib}=\theta_{ib}\sigma^2_{m}$ and $\sigma^2_{ie_j}=\theta_{ie_j}\sigma^2_{e}$, we have
\begin{equation}
E(t) = \sum\limits_{n = 1}^{{2^N} - 1} {{{( - 1)}^{|{E_n}| + 1}}{(\sum\limits_{j \in {E_n}} {\frac{{{N_{{e_j}}}{\theta _{ib}}}}{{{N_b}{\theta _{i{e_j}}}}}} )^{ - 1}}}  \cdot \frac{1}{{{\lambda _{me}}}},\tag{A.3}\label{A.3}
\end{equation}
where $\lambda_{me}=\sigma^2_{m}/\sigma^2_{e}$. One can observe from
Eq. (A.3) that $E(t)$ approaches to zero for $\lambda_{me} \to
\infty$. Meanwhile, using Eq. (43), we can obtain the mean of $t^2$
as Eq. (A.4) at the top of the following page,
\begin{figure*}
\begin{equation}
\begin{split}
E({t^2}) &= \int_0^\infty  {\frac{{N_b^2{x^2}}}{{\sigma _{ib}^4}}\sum\limits_{n = 1}^{{2^N} - 1} {{{( - 1)}^{|{E_n}| + 1}}\sum\limits_{j \in {E_n}} {(\frac{{{N_{{e_j}}}}}{{\sigma _{i{e_j}}^2}})} \exp ( - \sum\limits_{j \in {E_n}} {\frac{{{N_{{e_j}}}x}}{{\sigma _{i{e_j}}^2}}} )} dx} \\
&= \sum\limits_{n = 1}^{{2^N} - 1} {{{( - 1)}^{|{E_n}| + 1}}} \int_0^\infty  {\frac{{N_b^2{x^2}}}{{\sigma _{ib}^4}}\sum\limits_{j \in {E_n}} {(\frac{{{N_{{e_j}}}}}{{\sigma _{i{e_j}}^2}})} \exp ( - \sum\limits_{j \in {E_n}} {\frac{{{N_{{e_j}}}x}}{{\sigma _{i{e_j}}^2}}} )dx} \\
&= 2\sum\limits_{n = 1}^{{2^N} - 1} {{{( - 1)}^{|{E_n}| + 1}}} (\sum\limits_{j \in {E_n}} {\frac{{{N_{{e_j}}}\sigma _{ib}^2}}{{{N_b}\sigma _{i{e_j}}^2}}{)^{ - 2}}} \\
&= 2\sum\limits_{n = 1}^{{2^N} - 1} {{{( - 1)}^{|{E_n}| + 1}}{(\sum\limits_{j \in {E_n}} {\frac{{{N_{{e_j}}}{\theta _{ib}}}}{{{N_b}{\theta _{i{e_j}}}}}} )^{ - 2}}}  \cdot {(\frac{1}{{{\lambda _{me}}}})^2},
\end{split}\tag{A.4}\label{A.4}
\end{equation}
\end{figure*}
which shows that $E(t^2)$ approaches to zero for $\lambda_{me} \to \infty$. Since both $E(t)$ and $E(t^2)$ approach to zero as $\lambda_{me} \to \infty$, we can easily obtain that random variable $t$ approaches to zero with probability one for $\lambda_{me} \to \infty$. Hence, considering $t \to 0$ with probability one and using Taylor series expansion, we obtain
\begin{equation}
\exp (-t) = 1 -t + O(\frac{1}{{{\lambda _{me}}}}),\tag{A.5}\label{A.5}
\end{equation}
for $\lambda_{me} \to \infty$, where $O(\frac{1}{{{\lambda _{me}}}})$ represents high-order infinitesimals. Substituting $t = \frac{{{N_b}x}}{{\sigma _{ib}^2}}$ into Eq. (A.5) gives
\begin{equation}\nonumber
\exp ( - \frac{{{N_b}x}}{{\sigma _{ib}^2}}) = 1 - \frac{{{N_b}x}}{{\sigma _{ib}^2}} + O(\frac{1}{{{\lambda _{me}}}}),
\end{equation}
which in turn leads to
\begin{equation}
1-\exp ( - \frac{{{N_b}x}}{{\sigma _{ib}^2}}) = \frac{{{N_b}x}}{{\sigma _{ib}^2}} + O(\frac{1}{{{\lambda _{me}}}}),\tag{A.6}\label{A.6}
\end{equation}
which completes the proof of Eq. (45).

\begin{IEEEbiography}[{\includegraphics[width=1in,height=1.25in]{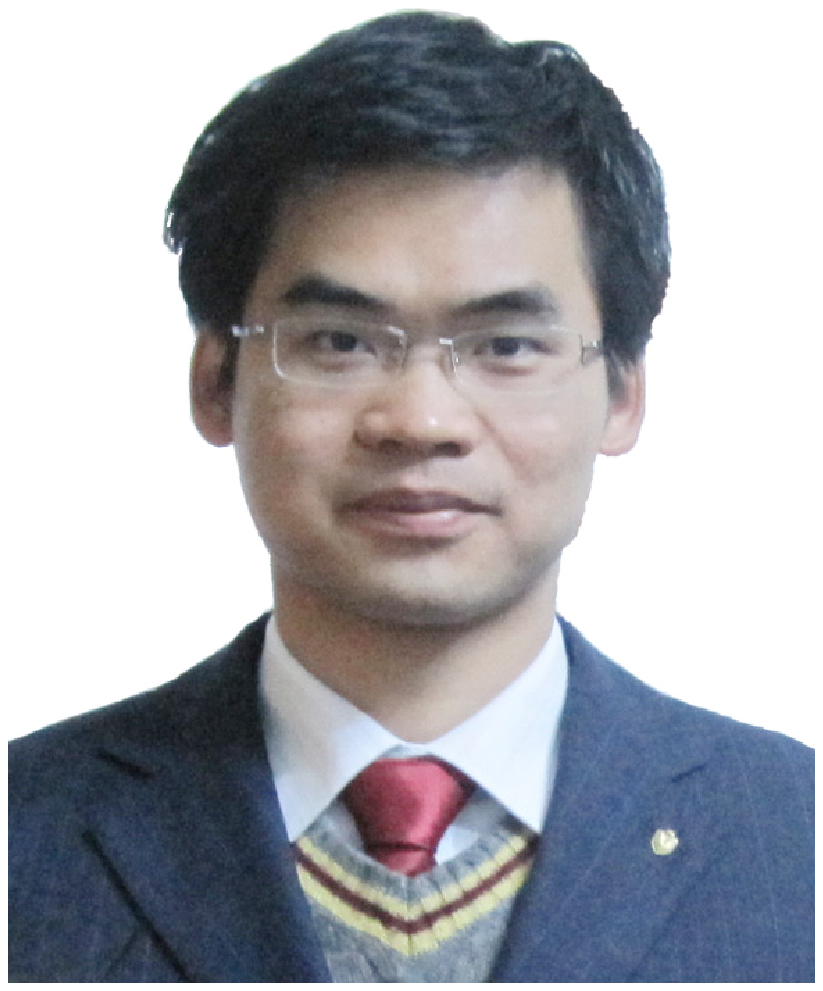}}]{Yulong
Zou} (S'07-M'12-SM'13) received the B.Eng. degree in information
engineering from the Nanjing University of Posts and
Telecommunications (NUPT), Nanjing, China, in July 2006, the first Ph.D. degree from the Stevens Institute of Technology, New Jersey, United States, in May 2012, and the second Ph.D. degree from NUPT, Nanjing, China, in July 2012.

Dr. Zou is currently serving as an editor for the IEEE
Communications Surveys \& Tutorials, IEEE Communications Letters,
EURASIP Journal on Advances in Signal Processing, and KSII
Transactions on Internet and Information Systems. He is also serving
as the lead guest editor for a special issue on ``Security Challenges and Issues in Cognitive Radio Networks" in the EURASIP Journal on
Advances in Signal Processing. In addition, he has acted as
symposium chairs, session chairs, and TPC members for a number of
IEEE sponsored conferences including the IEEE Wireless
Communications and Networking Conference (WCNC), IEEE Global
Telecommunications Conference (GLOBECOM), IEEE International
Conference on Communications (ICC), IEEE Vehicular Technology
Conference (VTC), International Conference on Communications in
China (ICCC), and so on.

His research interests span a wide range of topics in wireless
communications and signal processing including the cooperative
communications, cognitive radio, wireless security, and green
communications. In these areas, he has published extensively in
internationally renowned journals including the IEEE Transactions on
Signal Processing, IEEE Transactions on Communications, IEEE Journal
on Selected Areas in Communications, IEEE Transactions on Wireless
Communications, and IEEE Communications Magazine.

\end{IEEEbiography}

\begin{IEEEbiography}[{\includegraphics[width=1in,height=1.25in]{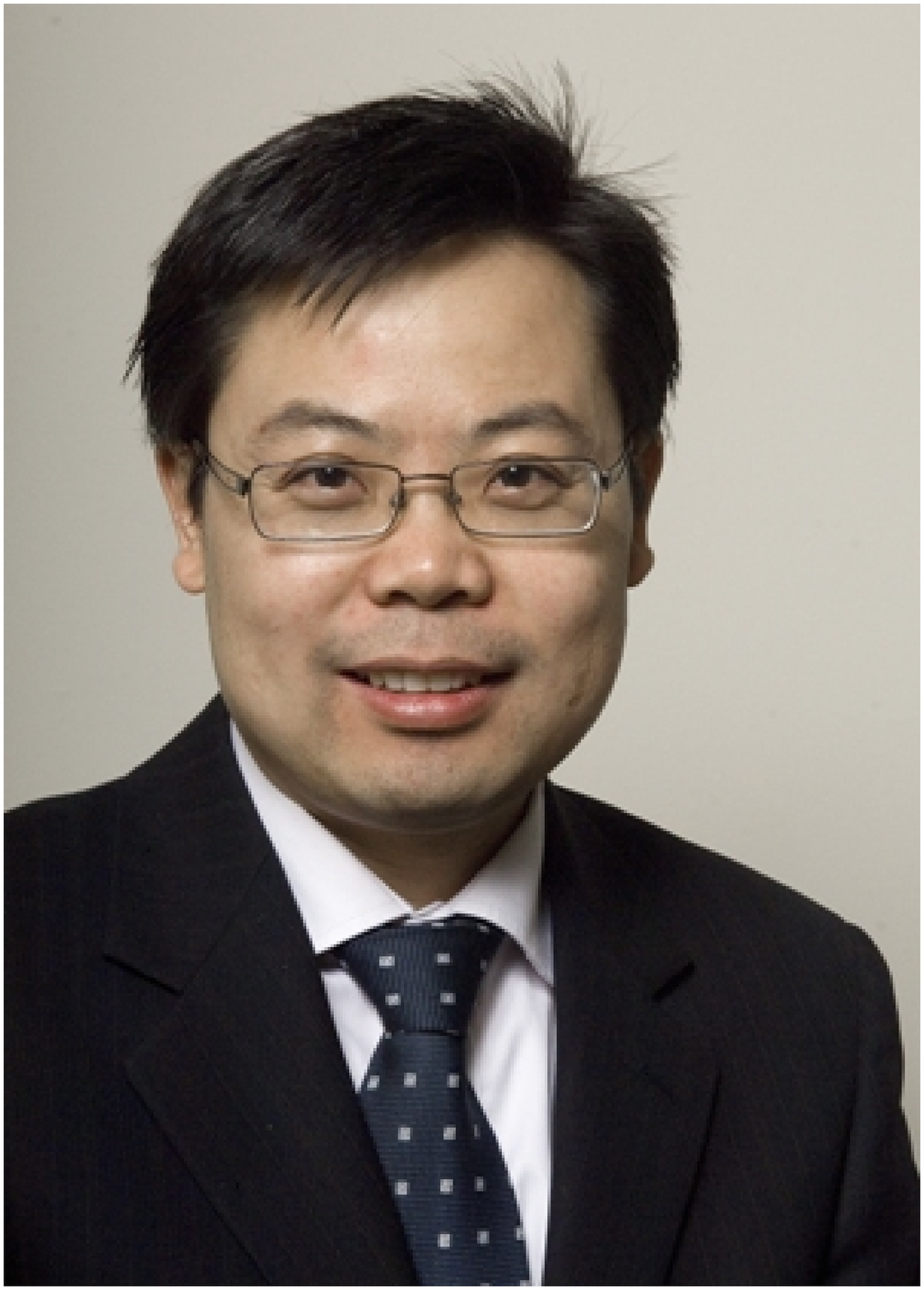}}]
{Xianbin Wang}(S'98-M'99-SM'06) is an Associate Professor at The
University of Western Ontario and a Canada Research Chair in
Wireless Communications. He received his Ph.D. degree in electrical
and computer engineering from National University of Singapore in
2001.

Prior to joining Western, he was with Communications Research Centre
Canada as Research Scientist/Senior Research Scientist between July
2002 and Dec. 2007. From Jan. 2001 to July 2002, he was a system
designer at STMicroelectronics, where he was responsible for system
design for DSL and Gigabit Ethernet chipsets. He was with Institute
for Infocomm Research, Singapore (formerly known as Centre for
Wireless Communications), as a Senior R \& D engineer in 2000. His
primary research area is wireless communications and related
applications, including adaptive communications, wireless security,
and wireless infrastructure based position location. Dr. Wang has
over 150 peer-reviewed journal and conference papers on various
communication system design issues, in addition to 23 granted and
pending patents and several standard contributions.

Dr. Wang is an IEEE Distinguished Lecturer and a Senior Member of
IEEE. He was the recipient of three IEEE Best Paper Awards. He
currently serves as an Associate Editor for IEEE Wireless
Communications Letters, IEEE Transactions on Vehicular Technology
and IEEE Transactions on Broadcasting. He was also an editor for
IEEE Transactions on Wireless Communications between 2007 and 2011.
Dr. Wang was involved in a number of IEEE conferences including
GLOBECOM, ICC, WCNC, VTC, and ICME, on different roles such as
symposium chair, track chair, TPC and session chair.

\end{IEEEbiography}

\begin{IEEEbiography}[{\includegraphics[width=1in,height=1.25in]{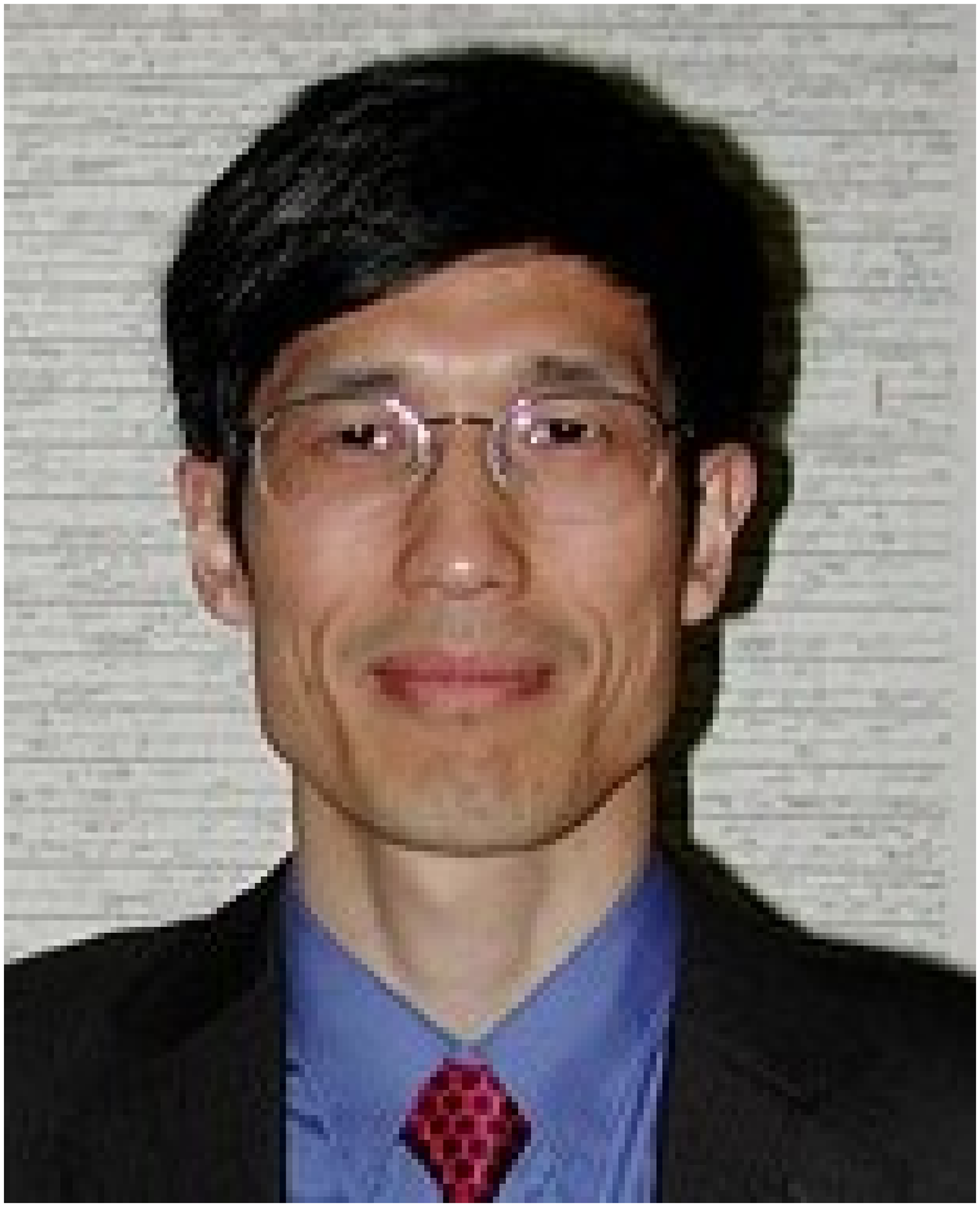}}]{Weiming Shen} is a Senior Research Scientist at the National Research Council Canada and an Adjunct Professor at Tongji University, China, and University of Western Ontario, Canada. He is a Fellow of IEEE. He received his Bachelor and Master¡¯s degrees from Northern (Beijing) Jiaotong University, China and his PhD degree from the University of Technology of Compi¨¨gne, France. His recent research interest includes agent-based collaboration technology and applications, wireless sensor networks. He has published several books and over 300 papers in scientific journals and international conferences in the related areas. His work has been cited over 6,000 times with an h-index of 37. He has been invited to provide over 60 invited lectures/seminars at different academic and research institutions over the world and keynote presentations / tutorials at various international conferences. He is a member of the Steering Committee for the IEEE Transactions on Affective Computing and an Associate Editor or Editorial Board Member of ten international journals (including IEEE Transactions on Automation Science and Engineering, Computers in Industry; Advanced Engineering Informatics; Service Oriented Computing and Applications) and served as guest editor for several other international journals.
\end{IEEEbiography}


\begin{thebibliography}{11}

\bibitem{IEEEhowto:1}
J. Mitola and G. Q. Maguire, \textquotedblleft Cognitive radio:
Making software radios more personal," \emph{IEEE Personal Commun.},
vol. 6, pp. 13-18, 1999.

\bibitem{IEEEhowto:2}
IEEE 802.22 Working Group, \textquotedblleft IEEE P802.22/D1.0 draft
standard for wireless regional area networks part 22: Cognitive
wireless RAN medium access control (MAC) and physical layer (PHY)
specifications: Policies and procedures for operation in the TV
bands," Apr. 2008.

\bibitem{IEEEhowto:3}
Y. Zou, Y.-D. Yao, and B. Zheng, \textquotedblleft Cooperative relay
techniques for cognitive radio systems: Spectrum sensing and
secondary user transmissions," \emph{IEEE Commun. Mag.}, vol. 50,
no. 4, pp. 98-103. Apr. 2012.

\bibitem{IEEEhowto:4}
Y. Zou, Y.-D. Yao, and B. Zheng, ``Cognitive transmissions with
multiple relays in cognitive radio networks," \emph{IEEE Trans.
Wireless Commun.}, vol. 10, no. 2, pp. 648-659, Feb. 2011.

\bibitem{IEEEhowto:5}
H. Li, ``Cooperative spectrum sensing via belief propagation in
spectrum-heterogeneous cognitive radio systems," in \emph{Proc. 2010
IEEE Wireless Commun. and Netw. Conf. (WCNC)}, Sydney Australia,
Apr. 2010.

\bibitem{IEEEhowto:6}
Y. Zou, J. Zhu, B. Zheng, and Y.-D. Yao, ``An adaptive cooperation
diversity scheme with best-relay selection in cognitive radio
networks," \emph{IEEE Trans. Signal Process.}, vol. 58, no. 10, pp.
5438-5445, Oct. 2010.

\bibitem{IEEEhowto:7}
R. Zhang and Y.-C. Liang, ``Exploiting multi-antennas for
opportunistic spectrum sharing in cognitive radio networks,"
\emph{IEEE J. Sel. Topics Signal Process.}, vol. 2, no. 1, pp.
88-102, Feb. 2008.

\bibitem{IEEEhowto:8}
L. Lu, X. Zhou, U. Onunkwo, and Y. Li, \textquotedblleft Ten years
of research in spectrum sensing and sharing in cognitive radio,"
\emph{EURASIP Journal on Wireless Commun. and Net.}, vol. 2012, DOI:
10.1186/1687-1499-2012-28, Jan. 2012.

\bibitem{IEEEhowto:9}
H. Li and Z. Han, ``Dogfight in spectrum: Combating primary user
emulation attacks in cognitive radio systems, part I: Known channel
statistics," \emph{IEEE Trans. Wireless Commun.}, vol. 9, no. 11,
pp. 3566-3577, Nov. 2010.


\bibitem{IEEEhowto:10}
T. Brown and A. Sethi, \textquotedblleft Potential cognitive radio
denial-of-service vulnerabilities and protection countermeasures: a
multi-dimensional analysis and assessment," in \emph{Proc. IEEE
International Conf. Cognitive Radio Oriented Wireless Net. Commun.},
Orlando Florida, Aug. 2007.

\bibitem{IEEEhowto:11}
C. E. Shannon, ``Communication theory of secrecy systems,"
\emph{Bell System Technical Journal}, vol. 28, pp. 656-715, 1949.

\bibitem{IEEEhowto:12}
A. D.Wyner, ``The wire-tap channel," \emph{Bell System Technical
Journal}, vol. 54, no. 8, pp. 1355-1387, 1975.

\bibitem{IEEEhowto:13}
S. K. Leung-Yan-Cheong and M. E. Hellman, ``The Gaussian wiretap
channel," \emph{IEEE Trans. Inf. Theory}, vol. 24, pp. 451-456, Jul.
1978.

\bibitem{IEEEhowto:14}
Z. Li, W. Trappe, and R. Yates, ``Secret communication via
multi-antenna transmission," in \emph{Proc. 41st Conf. Information
Sciences Systems}, Baltimore, MD, Mar. 2007.

\bibitem{IEEEhowto:15}
A. Khisti, G. Womell, A. Wiesel, and Y. Eldar, ``On the Gaussian
MIMO wiretap channel," in \emph{Proc. IEEE Int. Symp. Inf. Theory},
Nice, France, Jun. 2007.

\bibitem{IEEEhowto:16}
F. Oggier and B. Hassibi, ``The secrecy capacity of the MIMO wiretap
channel," \emph{IEEE Trans. Inf. Theory}, vol. 57, no. 8, pp.
4961-4972, Oct. 2007.

\bibitem{IEEEhowto:17}
Y. Zou, X. Wang, and W. Shen, ``Optimal relay selection for
physical-layer security in cooperative wireless networks,"
\emph{IEEE J. Sel. Areas Commun.}, vol. 31, no. 10, pp. 2099-2111, Oct. 2013.

\bibitem{IEEEhowto:18}
Y. Zou, X. Wang, and W. Shen, ``Intercept probability analysis of
cooperative wireless networks with best relay selection in the
presence of eavesdropping attack," in \emph{Proc. 2013 IEEE Intern.
Conf. Commun. (IEEE ICC 2013)}, pp. 1-5, Budapest, Hungary, Jun.
2013.

\bibitem{IEEEhowto:19}
E. G. Larsson, ``On the combination of spatial diversity and
multiuser diversity," \emph{IEEE Commun. Lett.}, vol. 8, no. 8, pp.
517-519, Aug. 2004.

\bibitem{IEEEhowto:20}
Q. H. Spencer, C. B. Peel, A. L. Swindlehurst, and M. Haardt, ``An
introduction to the multi-user MIMO downlink," \emph{IEEE Commun.
Mag.}, vol. 42, no. 10, pp. 60-67, Oct. 2004.

\bibitem{IEEEhowto:21}
S. Goel and R. Negi, ``Guaranteeing
secrecy using artificial noise," \emph{IEEE Trans. Wireless
Commun.}, vol. 7, no. 6, pp. 2180-2189, Jul. 2008.

\bibitem{IEEEhowto:22}
X. Zhou and M. McKay, ``Secure
transmission with artificial noise over fading channels: Achievable
rate and optimal power allocation," \emph{IEEE Trans. Veh. Tech.},
vol. 59, no. 8, pp. 3831-3842, Aug. 2010.

\bibitem{IEEEhowto:23}
D. Cabric, A. Tkachenko, and R. W. Brodersen, ``Experimental study
of spectrum sensing based on energy detection and network
cooperation," in \emph{Proc. First Int. Workshop Tech. and Policy
Access. Spectrum}, Boston, MA, Aug. 2006.

\bibitem{IEEEhowto:24}
G. Bansal, M. Hossain, and V. Bhargava, ``Optimal and suboptimal
power allocation schemes for OFDM-based cognitive radio systems,"
\emph{IEEE Trans. Wireless Commun.}, vol. 7, no. 11, pp. 4710-4718,
Nov. 2008.

\bibitem{IEEEhowto:25}
{M. Bloch, J. O. Barros, M. R. D.
Rodrigues, and S. W. McLaughlin, ``Wireless information-theoretic
security," \emph{IEEE Trans. Inf. Theory}, vol. 54, no. 6, pp.
2515-2534, Jun. 2008. }


\bibitem{IEEEhowto:26}
{L. Dong, Z. Han, A. P. Petropulu, and H. V. Poor, ``Improving
wireless physical layer security via cooperating relays,"
\emph{IEEE Trans. Signal Process.}, vol. 58, no. 3, pp. 1875-1888, Mar. 2010.}

\bibitem{IEEEhowto:27}
A. Mukherjee and A. Swindlehurst, ``Robust beamforming for security in MIMO wiretap channels with imperfect CSI," \emph{IEEE Trans. Signal Process.}, vol. 59, no. 1, pp. 351-361, Jan. 2011.


\bibitem{IEEEhowto:28}
H. Qin, \emph{et al.}, ``Optimal power allocation for joint beamforming and artificial noise design in secure wireless communications," in \emph{Proc. The 2011 IEEE Intern. Conf. Commun.}, Kyoto, Japan, June 2011.

\bibitem{IEEEhowto:29}
L. Zheng and D. Tse, ``Diversity and multiplexing: A fundamental
tradeoff in multiple antenna channels," \emph{IEEE Trans. Inform.
Theory}, vol. 49, no. 5, pp. 1073-1096, May 2003.

\bibitem{IEEEhowto:30}
{H. J. Kushner and P. A. Whiting, ``Convergence of
proportional-fair sharing algorithms under general conditions,"
\emph{IEEE Trans. Wireless Commun.}, vol. 3, no. 4, pp. 1250-1259, Jul. 2004.}

\end{thebibliography}
\end{document}